\documentclass[%
reprint,
superscriptaddress,
amsmath,amssymb,
aps,
prapplied,
]{revtex4-2}
\usepackage[utf8]{inputenc}
\usepackage[english]{babel}
\usepackage{blindtext}
\usepackage{graphicx}   
\graphicspath{{figs/}}   
\usepackage{dcolumn}
\usepackage{bm}
\usepackage[hidelinks]{hyperref}
\usepackage{epstopdf}
\usepackage{braket}
\usepackage{upgreek}
\usepackage{amsmath}
\usepackage{siunitx}
\usepackage{changes}
\usepackage{natbib}
\usepackage[title]{appendix}

\usepackage{multirow}

\begin{document}

\title{Superconducting microwave magnetometer for absolute flux detection}

\author{Simon Günzler}
\email{both authors contributed equally}
\affiliation{Institute for Quantum Materials and Technology, Karlsruher Institute of Technology, 76344 Eggenstein-Leopoldshafen, Germany}

\author{Patrick Winkel}
\email{both authors contributed equally}
\affiliation{Institute for Quantum Materials and Technology, Karlsruher Institute of Technology, 76344 Eggenstein-Leopoldshafen, Germany}

\author{Dennis Rieger}
\affiliation{Physikalisches Institut, Karlsruhe Institute of Technology, 76131 Karlsruhe, Germany}

\author{Kiril Borisov}
\affiliation{Institute of Nanotechnology, Karlsruhe Institute of Technology, 76344 Eggenstein-Leopoldshafen, Germany}

\author{Martin Spiecker}
\affiliation{Physikalisches Institut, Karlsruhe Institute of Technology, 76131 Karlsruhe, Germany}

\author{Alexey V. Ustinov}
\affiliation{Institute for Quantum Materials and Technology, Karlsruher Institute of Technology, 76344 Eggenstein-Leopoldshafen, Germany}
\affiliation{Physikalisches Institut, Karlsruhe Institute of Technology, 76131 Karlsruhe, Germany}
\affiliation{Russian Quantum Center, National University of Science and Technology MISIS, 119049 Moscow, Russia}

\author{Ioan M. Pop}
\affiliation{Institute for Quantum Materials and Technology, Karlsruher Institute of Technology, 76344 Eggenstein-Leopoldshafen, Germany}
\affiliation{Physikalisches Institut, Karlsruhe Institute of Technology, 76131 Karlsruhe, Germany}
\affiliation{Institute of Nanotechnology, Karlsruhe Institute of Technology, 76344 Eggenstein-Leopoldshafen, Germany}

\author{Wolfgang Wernsdorfer}
\email{wolfgang.wernsdorfer@kit.edu}
\affiliation{Institute for Quantum Materials and Technology, Karlsruher Institute of Technology, 76344 Eggenstein-Leopoldshafen, Germany}
\affiliation{Physikalisches Institut, Karlsruhe Institute of Technology, 76131 Karlsruhe, Germany}
\affiliation{Institute of Nanotechnology, Karlsruhe Institute of Technology, 76344 Eggenstein-Leopoldshafen, Germany}
\affiliation{Institut Néel, CNRS and Université Joseph Fourier, Grenoble, France}

\date{\today}

\begin{abstract}
Superconducting quantum interference devices (SQUIDs) are among the most sensitive detectors for out-of-plane magnetic field components. However, due to their periodic response with short modulation period $M = 1 \Phi_0$, determined by the magnetic flux quantum $\Phi_0 \approx 2.068\times 10^{-15}\,\mathrm{Wb}$, it is difficult to infer the value of the magnetic flux unambiguously, especially in case the magnetic flux enclosed in the SQUID loop changes by many flux quanta. Here, we demonstrate that by introducing a second degree of freedom in the form of a second SQUID, we substantially enhance the modulation period $M$ of our device without sacrificing sensitivity. As a proof of concept, we implement our idea by embedding two asymmetric direct current SQUIDs into a common tank circuit. By measuring the reflection coefficient of the device, we extract the two lowest energy eigenfrequencies as a function of the external magnetic flux created by a superconducting field coil, from which we experimentally deduce a modulation period $M \geq 15 \Phi_0$, as well as the magnetic offset-field $B_0 = 22\,\mathrm{nT}$ present in our experiment.
\end{abstract}


\maketitle

\section{Introduction}

Superconducting electronics (SCEs) combine the concept of integrated circuitry adapted from semiconductor industry, with the unique properties of superconducting thin films: low energy losses and intrinsic non-linearities. Operated below the critical temperature $T_\mathrm{c}$ of the superconductor, the large versatility of SCEs arises from the Josephson effect \cite{Jos62} and the development of superconducting quantum interference devices (SQUIDs), enabling applications in quantum information processing \cite{Nakamura1999,Devoret04,Clarke2008,Schoelkopf2008,Krantz19}, quantum hybrid systems \cite{Kroll18, Fornieri19,Pita-vidal19}, and quantum sensing \cite{Zhu2011,Kubo12,Schneider18,Serniak19,Scarlino2019}. 

Even well before the implementation of the first superconducting quantum bit \cite{Nakamura1999}, the high sensitivity of SQUIDs \cite{Clarke2006,Cleuziou2006} has been used for decades to measure many different physical quantities with unmatched precision, by encoding their dynamics into a change of the local magnetic flux experienced by the SQUID, ranging from magnetic stray fields of geological formations \cite{Clarke83}, nanoparticles \cite{Wernsdorfer1995,Wernsdorfer1999_Science} and molecular magnets \cite{Wernsdorfer2002}, to electrical currents in solid-state \cite{Kamper1971,Giffard1972} and even biological systems \cite{Grossman129,Cohen70,Cohen72,Romani82,Sternickel2006,K_rber_2016,Enpuku2017}. Owing to the interplay between the magnetic flux quantization in a superconducting loop \cite{Deaver61,Doll61} and the Josephson effect \cite{Jos62}, the SQUID response to an external magnetic field oriented perpendicular to the loop is a periodic function of the magnetic flux enclosed in the loop. The resulting high sensitivity is attributed to the small modulation period $M = \Phi_0$, which is determined by the magnetic flux quantum $\Phi_0 = h / (2 e) \approx 2.068\times 10^{-15}\,\mathrm{Wb}$, where $h$ is Planck's constant, and $e$ is the elementary charge. However, the very same periodicity prevents an unambiguous measurement of the absolute magnetic flux and field beyond a single flux quantum, without the use of fast feedback loops \cite{Drung2007}. Moreover, it is impossible to unambiguously determine the absolute magnetic offset field present using a single SQUID only.

In this work, we demonstrate that by using the combined response of two dc SQUIDs, differing in their loop surface area only, we can enhance the modulation period $M$ of the magnetic flux periodic response to arbitrary values, well beyond a single magnetic flux quantum. Because the flux dependence within a period is distinctive, we can determine the magnitude of external bias fields perpendicular to the SQUID plane directly from the device response. Since the response is symmetric around effective zero-flux, and the modulation period $M$ is large by design, we can detect the presence of magnetic offset-fields with high accuracy at the same time. For an absolute magnetic field calibration, the only uncertainty originates from the effective loop surface areas of the SQUIDs. 

\begin{figure*}[!t]
\begin{center}
\includegraphics[width = 2\columnwidth]{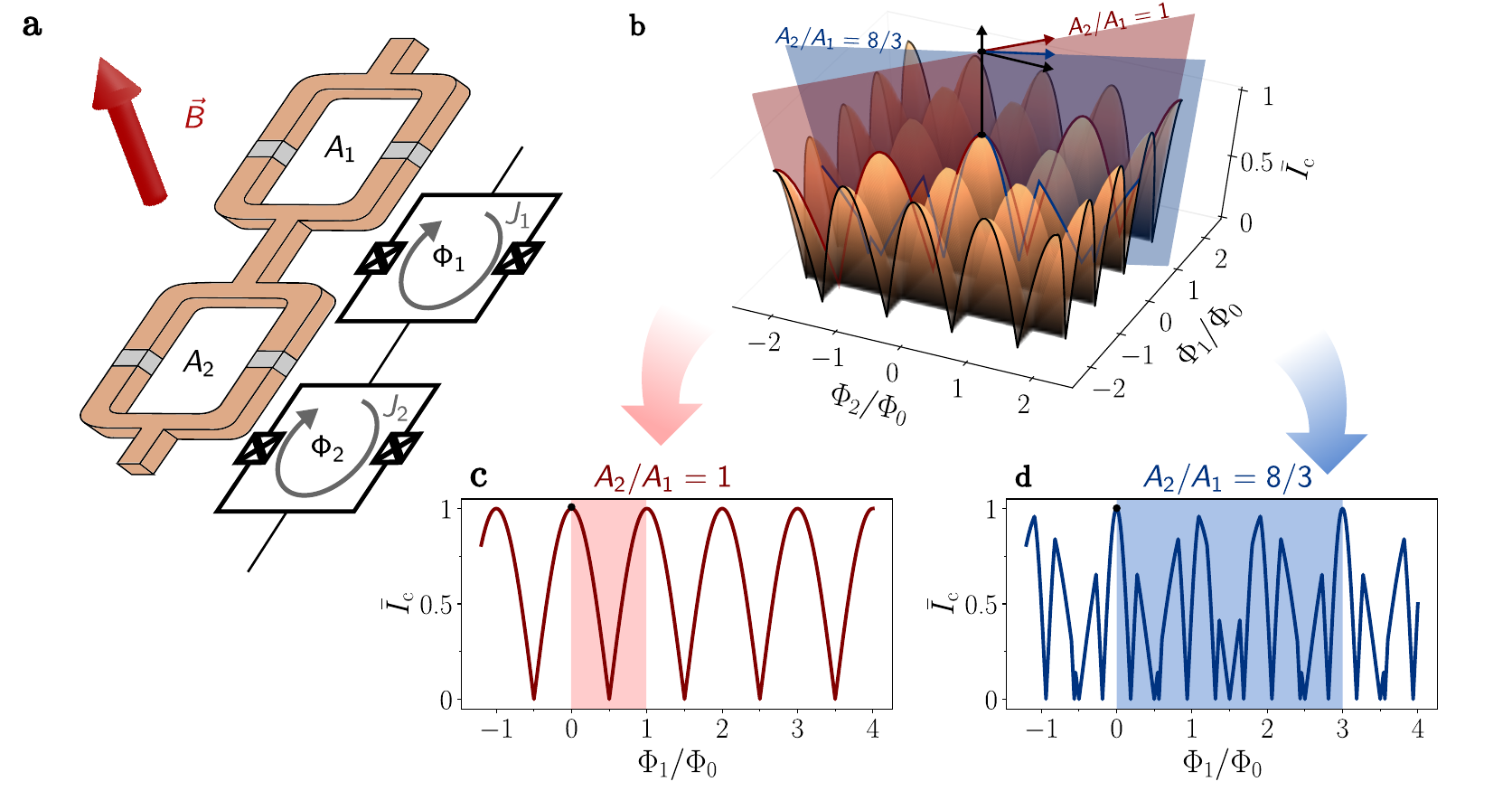}
\caption{\textbf{Magnetic flux modulation period engineering.} \textbf{a)} Artistic illustration and circuit diagram of two direct current (dc) superconducting quantum interference devices (SQUIDs) connected in series. Each SQUID consists of two identical Josephson junctions (JJ) with critical current $I_\mathrm{c}$ and Josephson capacitance $C_\mathrm{J}$ (framed cross symbols), embedded in parallel into superconducting loops with surface areas $A_1$ and $A_2$, respectively. The presence of an external magnetic field $\vec{B}$ (red arrow) induces screening currents $J_1$ and $J_2$ in the loops, which alter the effective critical current of each SQUID. \textbf{b)} Normalized, combined critical current as measured across both elements $\bar{I}_\mathrm{c} (\Phi_1 , \Phi_2) = \mathrm{min}\left\lbrace \bar{I}_\mathrm{c,1}(\Phi_1) , \bar{I}_\mathrm{c,2}(\Phi_2) \right\rbrace  / (2 I_\mathrm{c})$, depicted as a function of the magnetic flux $\Phi_i = \vec{B}_i \vec{A}_i$ $(i \in \{1,2\})$ enclosed in the SQUID loops. The periodicity along both axes is determined by the magnetic flux quantum $\Phi_0$. In case both SQUIDs are exposed to the same magnetic field ($\vec{B}_1 = \vec{B}_2$), the magnetic flux enclosed in the loops is related by their loop surface area ratio $\Phi_2 = r \Phi_1$, with $r = A_2 / A_1$. \textbf{c)} For the special case of integer values ($r \in \mathbb{Z}$), the modulation period $\bar{\Phi}$ of the combined response still remains a single flux quantum enclosed in the smaller loop, as illustrated for the trivial case $r = 1$ by the solid red line and the red shaded area. \textbf{d)} For non-integer values ($r \notin \mathbb{Z}$), the modulation period $M$ is enhanced, exemplified by the solid blue line for $r = 8 / 3$. For any rational number $r = a/b$, the modulation period is $M = b \Phi_0$, where $\Phi_0$ is the flux quantum enclosed in loop $A_1$. In practice, the achieved enhancement is limited by the experimental measurement resolution of the critical current.}
\label{fig1}
\end{center}
\end{figure*} 

We implement our concept by embedding two direct current (dc) SQUIDs into a common tank circuit, which is coupled to a 3D waveguide sample holder via its electrical dipole moment for readout. From resonance fluorescence measurements \cite{Astafiev10}, we extract the two lowest energy transitions of the system, denoted $f_+$ and $f_-$, which carry information on the magnetic flux enclosed in both SQUID loops. We confirm a modulation period beyond $M > 15 \, \Phi_0$ experimentally, with a theoretical limit of $M \sim 625 \Phi_0$, which we extrapolate from an effective circuit model. Further, we are able to determine the magnetic offset-field $B_{\perp,0} = 22\,\mathrm{nT}$ present in our magnetically shielded sample holder.

This article is organized as follows: In Sec.\,\ref{SEC:concept} we present the concept of the modulation period engineering, followed by the sample design and fabrication discussed in Sec.\,\ref{SEC:sample_desgin}. Sec.\,\ref{SEC:characterization} is devoted to the sample characterization in terms of the electromagnetic probe field, as well as the static and non-static external magnetic field. In Sec.\,\ref{SEC:conclusion} we conclude by summarizing the main results.

\section{Concept}
\label{SEC:concept}
\subsection{Modulation period engineering}
Conceptually, a direct-current (dc) SQUID consists of two JJs with critical currents denoted $I_\mathrm{c,1}$ and $I_\mathrm{c,2}$, respectively, which are embedded in parallel into a superconducting loop with surface area $A$. Due to the magnetic flux quantization observed in superconducting loops \cite{Deaver61,Doll61}, a circulating screening current $J$ is induced into the loop by an external magnetic field in case the enclosed flux $\Phi$ is not an integer multiple of the magnetic flux quantum $\Phi_0$. Notably, for the sake of completeness, there are scenarios in which this constrain is lifted, as shown in Ref.\,\cite{Friedrich19}. Because the screening current interferes with the unidirectional bias current flowing through both loop branches, the total critical current across the device $\bar{I}_\mathrm{c}$ is found to be a periodic function of the magnetic flux $\Phi$. In the simplest case, for which the critical currents of both JJs are identical $I_\mathrm{c,1} = I_\mathrm{c,2} = I_\mathrm{c}$, and the geometric inductance $L_\mathrm{g}$ attributed to the loop wire is negligible, the flux modulation of the SQUID's critical current $\bar{I}_\mathrm{c}$ is described by
\begin{equation}
\bar{I}_{\mathrm{c}} \left( \Phi \right) = 2 I_{\mathrm{c}}\left|\cos \left(\pi \frac{\Phi}{\Phi_0} \right)\right|,
\label{EQ:Ic_SQUID}
\end{equation}
and hence oscillates between its maximum value $2 I_\mathrm{c}$ and zero, when the enclosed flux $\Phi = A B_\perp$ is swept by means of an external magnetic field $B_\perp$ oriented out-of-plane with respect to the loop plane \cite{Tesche1977}. According to Eq.\,\ref{EQ:Ic_SQUID}, the modulation period $M$ of a single SQUID is equal to a single magnetic flux quantum $\Phi_0$. Since neighboring periods are indistinguishable due to the underlying symmetry of the response, the enclosed flux cannot be inferred unambiguously from the measured critical current. This limitation is avoidable by introducing a second degree of freedom in form of a second SQUID differing in its loop area. 

Figure\,\ref{fig1}a depicts an artistic illustration of two dc SQUIDs connected in series, together with a simplified circuit diagram assuming identical JJs (crossed boxes) but individual loop areas denoted $A_1$ and $A_2$, while neglecting the geometric loop inductance completely. For an arbitrary magnetic field $\vec{B}$ (red arrow), the magnetic flux $\Phi_i = A_i B_{\perp,i} $ with $i \in \{1,2\}$ enclosed in each SQUID loop depends on the field component $B_{\perp,i}$ oriented perpendicular to the SQUID plane, and the respective loop area $A_i$. Since the external bias current is identical for both SQUIDs, the critical current measured across both SQUIDs at the same time will be defined by the smaller of the two $\bar{I}_\mathrm{c} (\Phi_1 , \Phi_2) = \mathrm{min}\left\lbrace \bar{I}_\mathrm{c,1}(\Phi_1) , \bar{I}_\mathrm{c,2}(\Phi_2) \right\rbrace$, where the individual critical currents are determined by Eq.\,\ref{EQ:Ic_SQUID}. Figure\,\ref{fig1}b shows the normalized critical current $\bar{I}_\mathrm{c} (\Phi_1 , \Phi_2) / (2 I_\mathrm{c})$ as a function of the magnetic flux $\Phi_1$ and $\Phi_2$, illustrating the periodicity caused by the individual SQUIDs.  

Provided both SQUIDs are exposed to the same magnetic field, i.e.\,the external magnetic field is spatially homogeneous across both loops, the two enclosed fluxes are not independent degrees of freedom, but related by the loop area ratio $r = A_2 / A_1$, thus yielding $\Phi_2 = r \Phi_1$. For the special case of an integer loop area ratio $r \in \mathbb{Z}$, illustrated by way of example for $r =  1$ (left-hand panel), the modulation period of the combined response is still a single flux quantum enclosed in the SQUID loops (red shaded area). However in contrast, for any arbitrary rational but non-integer area ratio $r = a / b$ with $a,b \in \mathbb{Z}$, the period $M$ is enhanced and determined by the denominator of the reduced loop area ratio, as illustrated for $r = 8 / 3$ (right-hand panel, blue shaded area). Since the response is distinctive within a single period, any bias field inside the first period can be deduced unambiguously from the local response around the selected bias point. Conceptually, there is no limit for the enhancement of the modulation period as long as the critical current can be measured sufficiently accurate. In the following section, we will discuss a microwave readout implemented by embedding both SQUIDs in a common tank circuit, which potentially offers the following advantages: faster repetition, wireless readout, and entirely non-dissipative, which facilitates their embedding in a variety of superconducting and quantum hybrid quantum circuits.
 
\subsection{Microwave readout} 
The enhancement of the magnetic flux modulation period $M$, as discussed in the previous section (see Fig.\,\ref{fig1}), relies on the simultaneous readout of the flux enclosed in two dc SQUIDs with a non-integer loop area ratio $r \notin \mathbb{Z}$, and can be either inferred from transport or from microwave measurements. In this work, we focus on the latter case. 

From an electrical engineering perspective, a conventional JJ based on a tunneling contact between two superconducting electrodes is associated with a capacitance $C_\mathrm{J}$ and a non-linear kinetic inductance $L_\mathrm{k}$ in parallel, effectively forming an LC circuit. Since the kinetic inductance originates from the finite inertia experienced by the Cooper pairs due to the tunneling process, the linear part, commonly referred to as the Josephson inductance $L_\mathrm{J} = \Phi_0 / (2 \pi I_\mathrm{c})$, is linked to the electrical transparency of the JJ, i.e. the critical current $I_\mathrm{c}$. A similar relation is found for a dc SQUID, but with a critical current and Josephson inductance that depend on the magnetic flux $\Phi$ enclosed in the SQUID loop as discussed in the previous section. In the general case of a dc SQUID containing two potentially different size JJs, the Josephson inductance is 
\begin{equation}
L_\mathrm{J} (\Phi) = \frac{L_\mathrm{J,0}}{\left|\cos \left(\pi \frac{\Phi}{\Phi_0} \right)\right|\sqrt{1 + d^2 \tan \left(\pi \frac{\Phi}{\Phi_0} \right)^2}},
\label{EQ:LJ_SQUID}
\end{equation}
where $L_\mathrm{J,0} = \Phi_0 / (2 \pi [I_\mathrm{c,1} + I_\mathrm{c,2}])$ is the Josephson inductance in zero-field, and the parameter ${d = |I_{\mathrm{c},1} - I_{\mathrm{c},2}| / (I_{\mathrm{c},1} + I_{\mathrm{c},2})}$ accounts for a critical current asymmetry. As a consequence, by embedding a SQUID into a tank circuit, the information on the enclosed magnetic flux can be encoded in the resonance frequency of the circuit. 

The performance of such an implementation is highly dependent on the experimental ability to measure the resonance frequency of the circuit with high precision as fast as possible. For a given transition, the frequency resolution achieved is related to the spectral linewidth $\Gamma$, which can be the result of uncontrolled losses and the finite coupling to the measurement apparatus. A potential limitation for the speed of the readout can arise from the intrinsic non-linearity of the kinetic inductance. In a resonant circuit, the type of non-linearity introduced by a SQUID results in an anharmonic energy spectrum in the photon number basis, for which the frequency difference between the first two lowest energy transitions is captured by the anharmonicity $\alpha = 2 \pi \times (f_{12} - f_{01})$. Here, $f_{ij} = (E_j - E_i) / h$ is the transition frequency between the $i$-th and the $j$-th energy level. Provided the anharmonicity is significantly larger than the spectral linewidth $\Gamma$ of the transition ($\alpha \gg \Gamma$), the circuit is located in the quantum bit (qubit) regime \cite{Andersen2020}, which is clearly distinguishable from the opposite case ($\alpha \ll \Gamma$) typically found in most kinetic inductance detectors \cite{Valenti19}. In our case, the device described in the following sections is located in the qubit regime, rendering the readout particularly power dependent as we will discuss in Sec.\,\ref{SEC:power_calibration}. However, we would like to emphasize that this is not a requirement for the implementation of our concept.

\section{Sample design and fabrication}
\label{SEC:sample_desgin}
\begin{figure}[!t]
\begin{center}
\includegraphics[width = 1\columnwidth]{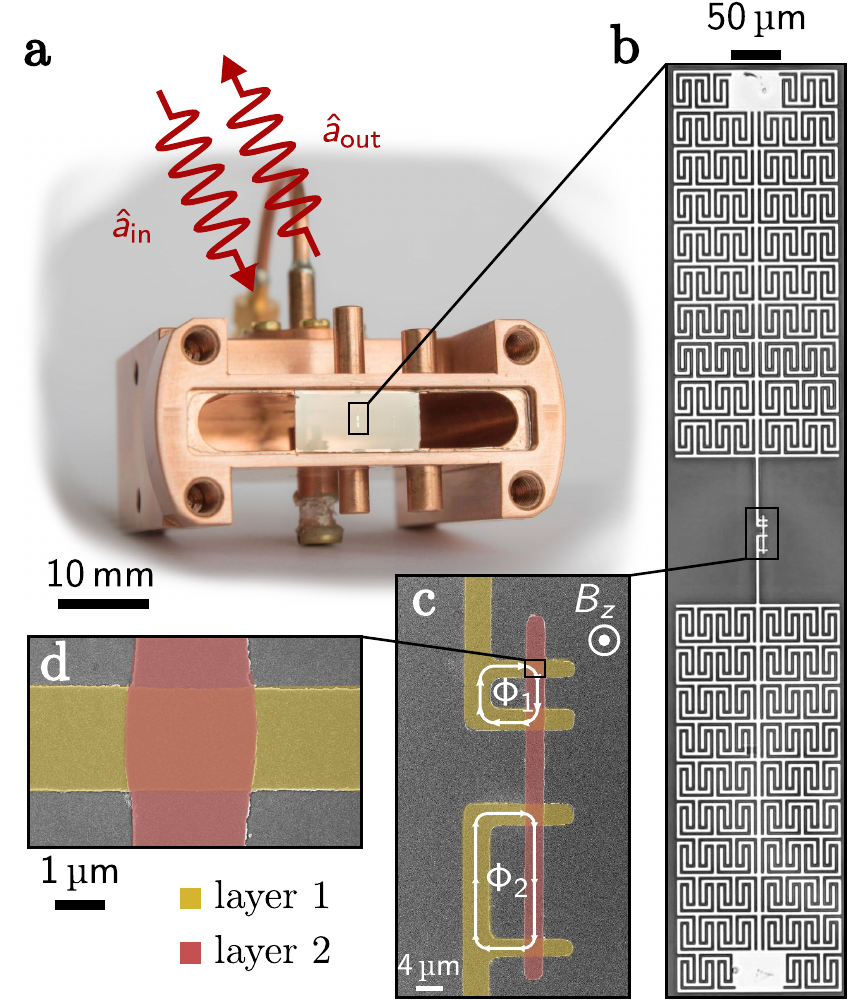}
\caption{\textbf{Device design.} \textbf{a)} Photograph of the 3D copper waveguide sample holder, conceptually similar to Ref.\,\cite{Kou18}, with the device mounted in the center. The sample holder is equipped with a single microwave port measured in reflection (top), where $\hat{a}_\mathrm{in}$ and $\hat{a}_\mathrm{out}$ are the single-mode field amplitudes of the incident and outgoing signal, respectively, and, optionally, with a 2D vector magnet (not shown, see Ref.\,\cite{Winkel_2020}). \textbf{b)} Optical microscopy image of the device consisting of two dc SQUIDs (center), galvanically coupled to a capacitor designed in the form of an antenna. The electrodes of the antenna are implemented in a fractal design to impede the formation of screening currents induced by external magnetic fields, and its electric dipole moment couples to the electric field of the sample holder. \textbf{c)} False-colored scanning electron microscopy (SEM) image of the two SQUIDs. The loop areas are $A_1 \approx 50\,\si{\micro\metre^2}$ (top) and $A_2 \approx 140\,\si{\micro\metre^2}$ (bottom), indicated by the closed white lines. The device fabrication is based on a two-step optical lithography process and zero-angle pure aluminum (Al) thin films. The overlap between the first (yellow) and second (light red) layer form the Josephson junctions (JJs), with overlap areas $A_\mathrm{JJ,1} \approx 6\pm0.5\,\si{\micro\metre^2}$ and $A_\mathrm{JJ,2} \approx 8\pm0.5\,\si{\micro\metre^2}$ (shown in \textbf{d}). By design, the two JJs in each SQUID are not identical to limit the frequency tunability of the device.}
\label{fig2}
\end{center}
\end{figure}

The sample design is composed of two SQUIDs which share a common in-plane capacitance $C_\mathrm{s}$ in the shape of a microwave antenna, forming a circuit with two distinct, magnetic field dependent frequencies $f_+ (\Phi_1 , \Phi_2)$ and $f_- (\Phi_1 , \Phi_2)$. For readout, the electric dipole moment of the antenna couples both modes simultaneously to the electric field of a 3D waveguide sample holder (see Fig.\,\ref{fig2}a) \cite{Kou18,Winkel_2020}. The coupling strength determines the spectral linewidth of the modes. In order to reduce the detrimental effect of persistent currents in the antenna pads on the device performance, induced by magnetic fields applied out-of-plane, the antenna is implemented in a fractal design (see Fig.\,\ref{fig2}b). 

The two SQUIDs are designed identical in terms of their Josephson inductance, but are both asymmetric with respect to the critical currents of their individual JJs, denoted $I_{\mathrm{c},i1}$ and $I_{\mathrm{c},i2}$, where $i \in \{1,2\}$ is the SQUID index. This asymmetry is captured by an individual asymmetry parameter ${d_i = |I_{\mathrm{c},i1} - I_{\mathrm{c},i2}| / (I_{\mathrm{c},i1} + I_{\mathrm{c},i2})}$ for each SQUID, whose main implication is a reduction in the absolute frequency tunability of both modes (see App.\,\ref{ASEC:Device eigenmodes}). In our design, the junction asymmetry is $d_1 = d_2 = 0.14$ for both SQUIDs, since the lower cut-off frequency of our waveguide sample holder makes a measurement of our device below $4.5\,\mathrm{GHz}$ significantly slower. The only intentional difference between the SQUIDs are the loop areas, $A_1$ and $A_2$, for the reasons discussed in Sec.\,\ref{SEC:concept}. From scanning electron microscopy (SEM) images (see Fig.\,\ref{fig2}c), we extract $A_1 = 50\,\si{\micro \metre^2}$ and $A_2 = 140\,\si{\micro \metre^2}$, bounded by the circumference measured in the center of the loop wires (white solid lines), resulting in a loop area ratio $r = 14 / 5$ and an effective modulation period $M = 5\,\Phi_0$.

The sample is patterned in a standard two-step optical lithography process, with each step followed by a zero-angle evaporation of $30\,\mathrm{nm}$ and $40\,\mathrm{nm}$ pure aluminum (Al) thin films, respectively. The four JJs are formed by the overlap area between these two Al layers (see Fig.\,\ref{fig2}d). Before we deposit the second Al layer, we remove the native oxide from the first layer by means of an argon milling process \cite{Gruenhaupt17}, followed by a static oxidation for $45\,\mathrm{min}$ in a controlled oxygen atmosphere, with an oxygen partial pressure $p_\mathrm{O_2} = 20\,\mathrm{mbar}$. Thanks to the remarkable large overlap areas $A_\mathrm{JJ,1} \approx 6\pm0.5\,\si{\micro \metre^2}$ and $A_\mathrm{JJ,2} \approx 8\pm0.5\,\si{\micro \metre^2}$, the resulting junction capacitances $C_\mathrm{J,1} \approx 300\,\mathrm{fF}$ and $C_\mathrm{J,2} \approx 400\,\mathrm{fF}$ are almost an order of magnitude larger than the antenna capacitance $C_\mathrm{s} = 62\,\mathrm{fF}$, and, therefore dominate the charging energy $E_{\mathrm{c},\pm} = e^2 / (2 C_\pm)$ of both modes, where $C_\pm$ is the total capacitance. Since the charging energy is small compared to the Josephson energy $E_\mathrm{J,\pm}$, both modes are deep in the so-called transmon regime \cite{Koch07}, with an estimated anharmonicity $\alpha_\pm \approx 2\pi \times 25\,\mathrm{MHz}$ (see App.\,\ref{ASEC:Transmon_anharmonicity}).

\begin{figure*}[!t]
\begin{center}
\includegraphics[width = 2\columnwidth]{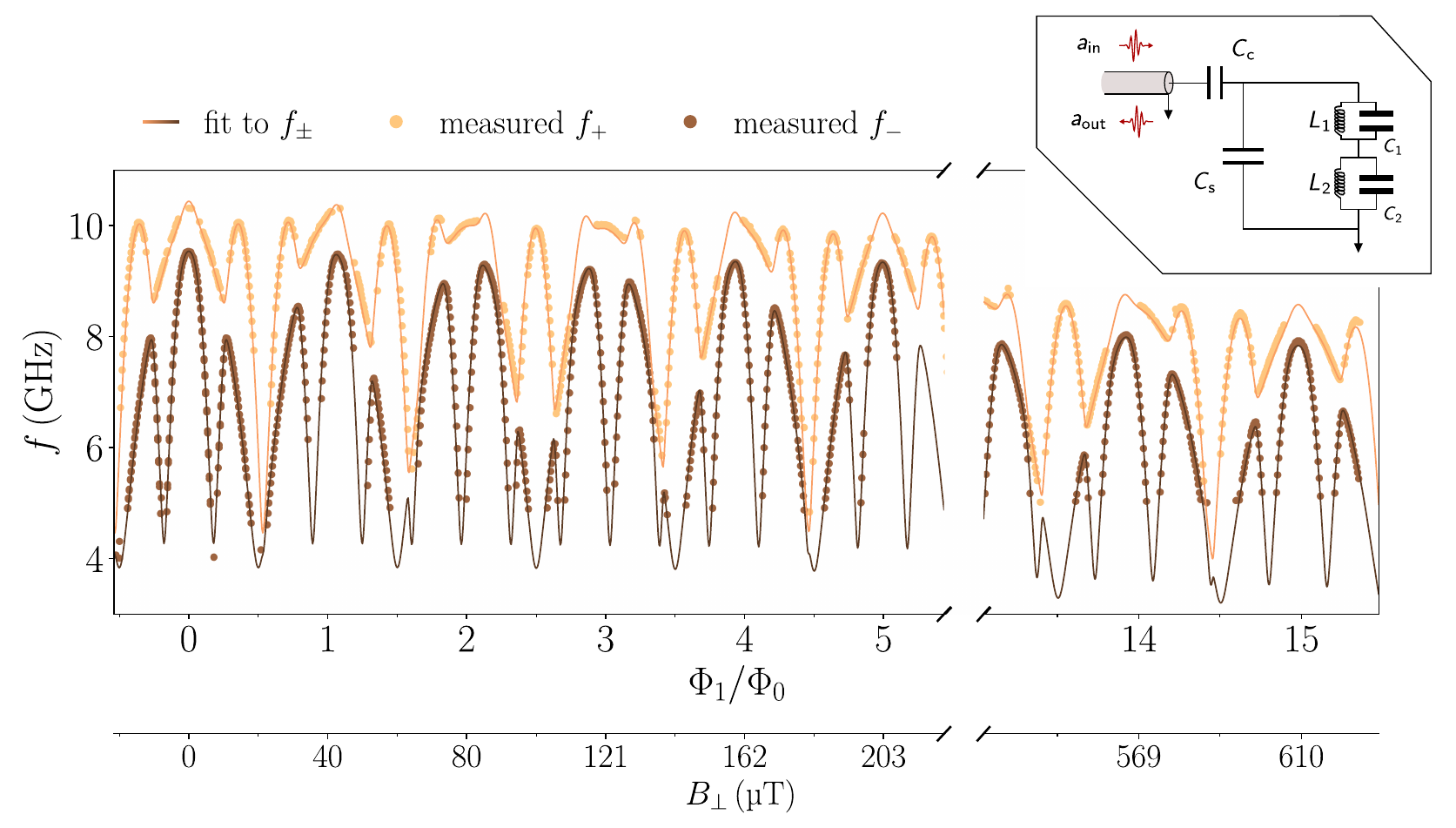}
\caption{\textbf{Magnetic flux modulation.} Measured eigenfrequencies $f_-$ (dark brown markers) and $f_+$ (light brown markers) of a device similar to Fig.\,\ref{fig2} as a function of the magnetic flux quanta $\Phi_1 / \Phi_0$ enclosed in the SQUID loop with smaller surface area $A_1$ (see Fig.\,\ref{fig2}c). The frequencies are extracted from continuous wave measurements of the reflection coefficient $S_{11} = a_\mathrm{out} / a_\mathrm{in}$ through the input port of the sample holder (see Fig.\,\ref{fig3_fluo}). Thanks to the non-integer surface area ratio $r = A_2 / A_1$, the combined flux modulation of both modes shows a significantly enhanced modulation period $M \gg \Phi_0$, exceeding the experimentally investigated field range. The dark and light solid lines illustrate the result of a simultaneous fit to both eigenmodes using a linearized circuit model with two degrees of freedom (see inset and App.\,\ref{ASEC:Device eigenmodes}), from which we extract the SQUID loop area ratio $r = 2.8048$, corresponding to a theoretical modulation period $M = 625 \, \Phi_0$. The remaining fit parameters of the model are the SQUID capacitances $C_1 = 722\,\mathrm{fF}$ and $C_2 = 718\,\mathrm{fF}$, the zero-field Josephson inductances $L_{\mathrm{J},1} = 322\,\mathrm{pH}$ and $L_{\mathrm{J},2} = 324\,\mathrm{pH}$, the total coupling capacitance $\tilde{C}_\mathrm{s} = C_\mathrm{s} + C_\mathrm{c} = 71\,\mathrm{fF}$, the SQUID asymmetry parameters $d_1 = 0.149$ and $d_2 = 0.184$, and the offset field $B_{\perp,0} = 22\,\mathrm{nT}$. The prediction of our model is in agreement with the measured data for the first few flux quanta. With increasing external magnetic flux, we observe a simultaneous lowering of both transition frequencies, which can be captured by the suppression of the superconducting gap parameter $\Delta_\mathrm{Al}$ of the pure aluminum thin film (see App.\,\ref{ASEC:Magnetic_flux_modulation}).}
\label{fig3}
\end{center}
\end{figure*}

\section{Sample characterization}
\label{SEC:characterization}
The key feature of our device is the magnetic field dependence of its two lowest energy eigenmodes, with the corresponding transition frequencies denoted $f_+$ and $f_-$. The indices $"  + "$ and $" - "$ indicate the experimentally accessible dressed basis representation, which accounts for the potential hybridization of both SQUID modes mediated by the shared antenna capacitor. In the experiment, we obtain the flux modulation of $f_+$ and $f_-$ by measuring the complex reflection coefficient $S_{11}$ in continuous wave, as a function of the probe frequency $f$ and the bias current $I_\mathrm{b}$ applied to a superconducting field coil, whose magnetic field is oriented perpendicular to the SQUID plane within machining precision, similar to Ref.\,\cite{Winkel_2020}. In this work, we extract the resonance frequencies $f_+$ and $f_-$ in post processing, but one can imagine a significantly faster readout and processing using a dedicated control hardware based on field programmable gate arrays \cite{Campagne_2013,Gebauer_2020}.  
 
In the presented implementation of our device shown in Fig.\,\ref{fig2}, the microwave response is not only distinctively nonlinear in terms of the applied magnetic field, but also in terms of the applied microwave drive power $P_\mathrm{in}$. Since this situation will have a direct consequence on the signal-to-noise ratio, we discuss the power dependence in context of the noise equivalent magnetic field.

\subsection{Magnetic flux calibration}
\label{SEC:flux_calibration}

Figure\,\ref{fig3} depicts the extracted frequencies $f_+$ (light brown markers) and $f_-$ (dark brown markers) as a function of the calibrated magnetic flux $\Phi_1$ enclosed in the smaller SQUID loop with surface area $A_1 = 50\,\si{\micro\metre}^2$. Following from the non-integer surface area ratio $r \notin \mathbb{Z}$ realized in our design, the flux modulation of both modes remains unique in a large range, enabling an unambiguous determination of the out-of-plane magnetic field. Moreover, since the flux modulation is only symmetric around effective zero-field, provided both SQUIDs experience the same magnetic field, the presence of a constant magnetic offset field $B_{\perp,0}$ is reflected in a positive or negative shift of the whole modulation pattern, depending on the orientation of the offset field. 

The solid lines indicate the fitting results to a linearized circuit model with two degrees of freedom, as illustrated in the top right corner of Fig.\,\ref{fig3}, from which we extract the electrical circuit parameters in zero-field (see App.\,\ref{ASEC:Device eigenmodes}), as well as the parameters relevant for the flux modulation, namely the global offset field $B_{\perp,0} = 22 \, \mathrm{nT}$ and the surface area ratio $r = 2.8048$ (see App.\,\ref{ASEC:Magnetic_flux_modulation}). Following the concept described in Fig.\,\ref{fig1}, we deduce $M = 625 \,\Phi_0$ from the extracted loop area ratio, which translates into a magnetic field $B_{\perp,M} = 25\,\mathrm{mT}$ using the loop area $A_1$. The measured value of $r$ is in quantitative agreement with the estimate taken from SEM images (see Fig.\,\ref{fig2}), and illustrates that the finite uncertainty introduced by the fabrication process enhances the modulation period compared to the design. The measured global offset field is roughly three orders of magnitude smaller than earth's magnetic field. We conclude that the small value of $B_{\perp,0}$ is due to the cylindrical $\si{\micro}$-metal shielding, surrounding our sample  similar to Ref.\,\cite{Gruenhaupt18}.

For the remaining parameters of our circuit model we find $C_1 = 722\,\mathrm{fF}$ and $C_2 = 718\,\mathrm{fF}$ for the SQUID capacitances, $L_{1} = 322\,\mathrm{pH}$ and $L_{2} = 324\,\mathrm{fF}$ for the zero-field Josephson inductances, $C_\mathrm{s} + C_\mathrm{c} = 71\,\mathrm{fF}$ for the total antenna capacitance, as well as $d_1 = 0.149$ and $d_2 = 0.184$ for the SQUID asymmetry parameters. These values are in good agreement with estimates deduced from SEM images of the JJs, and are discussed in App.\,\ref{ASEC:fitting_parameters} in greater detail. 

With increasing magnetic flux, we observe a decrease in the frequency modulation amplitude of both eigenmodes (see Fig.\,\ref{fig3} right-hand side), which can be caused by interference effects inside the JJs, or a suppression of the superconducting gap parameter $\Delta_\mathrm{Al}$ of the pure Al thin films \cite{Tin04}. Notably, we cannot distinguish between both effects within the measurement range shown in Fig.\,\ref{fig3}. While we expect both effects to be present at the same time, we can deduce a minimal critical magnetic field in the out-of-plane direction of $B_\mathrm{c,\perp} = 1.00\,\mathrm{mT}$, which limits the applicable range of our concept in the presented device, independent of the successful enhancement of the modulation period $M$. Moreover, at higher magnetic fields we observe jumps in the transition frequencies, probably caused by moving magnetic vortices trapped in the antenna pads.

\begin{figure}[!t]
\begin{center}
\includegraphics[width = 1\columnwidth]{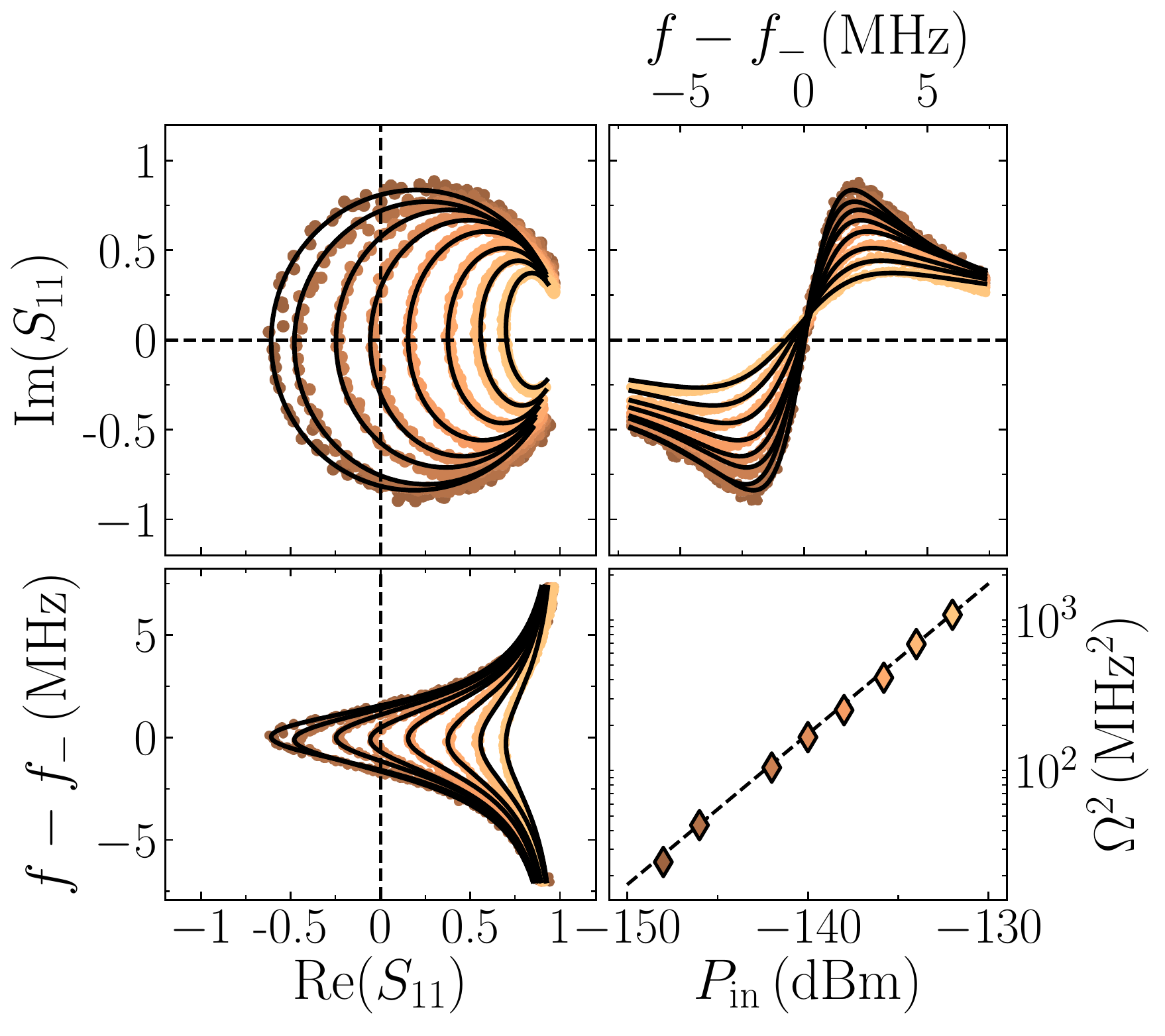}
\caption{\textbf{Power calibration.} Reflection coefficient $S_{11}$ measured as a function of frequency $f$ for different probe powers $P_\mathrm{in}$ in close vicinity to the transition frequency $f_- = 7.315\,\mathrm{GHz}$. With increasing power, the shape in the complex plane transits from a circular shape to an increasingly elliptic shape (top left), which is the signature of a non-linear system with distinct transition frequencies (qubit regime). From fits to the frequency dependence of the reflection coefficient (top right, bottom left) according to Eq.\,\ref{EQ:reflection_qubit}, indicated by the black solid lines, we characterize the mode with an external and internal decay rate $\kappa = 2 \pi \times 3.4\,\mathrm{MHz}$ and $\gamma = 2 \pi \times 0.5\,\mathrm{MHz}$, respectively, a pure dephasing rate $\Gamma_\phi = 2 \pi \times 2.0\,\mathrm{MHz}$, as well as a power dependent Rabi frequency $\Omega_\mathrm{R}$. The attenuation $A = 90\,\mathrm{dB}$ in the input line is deduced by comparing the expected and measured Rabi frequency (bottom right).}
\label{fig3_fluo}
\end{center}
\end{figure}

\subsection{Power calibration}
\label{SEC:power_calibration}

In contrast to a linear device, the distinct power dependence of the reflection coefficient discussed in this section gives rise to a non-monotonic relation between the signal-to-noise ratio and the applied readout power. Since the characteristic features of the modes vanish at high powers, there is an optimal readout power used to extract the transition frequencies from the measured reflection coefficient. 

For a sufficiently anharmonic energy spectrum ($\alpha \gg \Gamma$), the frequency and signal power dependence of the single-port complex reflection coefficient $S_{11}$ measured in frequency vicinity to a transition can be mapped onto the response of an effective two-level system \cite{Astafiev10,Winkel_2020}
\begin{equation}
S_{11} = 1 - \kappa \frac{\Gamma_1 (\Gamma_2 + i \Delta)}{\Gamma_1 (\Gamma_2^2 + \Delta^2) + \Gamma_2 \Omega_\mathrm{R}^2}.
\label{EQ:reflection_qubit}
\end{equation}
Here, $\kappa$ is the external coupling rate to the drive port, $\gamma$ is the internal decay rate, $\Delta = \omega_\mathrm{d} - \omega_0$ is the frequency detuning between drive frequency $\omega$ and transition frequency $\omega_0$, $\Gamma_1 = \kappa + \gamma$ is the total energy relaxation rate, $\Gamma_2$ is the dephasing rate, and $\Omega_\mathrm{R} (P_\mathrm{in})$ is the drive power dependent Rabi-frequency. In general, the dephasing rate $\Gamma_2 = \Gamma_1 / 2 + \Gamma_\phi$ is a combination of energy relaxation and pure dephasing at rate $\Gamma_\phi$ caused by time fluctuations of the transition frequency $\omega_0$. Even though both transitions of our device $f_+$ and $f_-$ are measured through the same input port, the frequency detuning between both is always large enough to approximate the response with Eq.\,\ref{EQ:reflection_qubit} in close vicinity to each transition. 

Figure\,\ref{fig3_fluo} depicts the reflection coefficient $S_{11}$ in the complex plane, as well as the frequency dependence of its real and imaginary part $\Re(S_{11})$ and $\Im(S_{11})$, respectively, measured around $f_- = 7.315\,\mathrm{GHz}$ for increasing probe powers $P_\mathrm{in}$. For the lowest probe power $P_\mathrm{in} = -148\,\mathrm{dBm}$ shown (dark brown markers), the reflection coefficient resembles a circle in the complex plane since the power dependent Rabi frequency is significantly smaller than the external coupling rate ($\Omega_\mathrm{R} \ll \kappa$). From the radius of the circle, we can infer the internal and external decay rates $\gamma$ and $\kappa$, respectively, which are representative for microwave losses into uncontrolled degrees of freedom (internal) and the input port (external). With increasing probe power, the reflection coefficient becomes increasingly elliptic, which is in agreement with the model of Eq.\,\ref{EQ:reflection_qubit} (solid black lines). From the fits to the data we can extract the resonance frequency $f_- = 7.315 \,\mathrm{GHz}$, the internal and external decay rates $\gamma = 2\pi \times 0.5 \, \mathrm{MHz}$ ($Q_\mathrm{i} = 14000$) and $\kappa = 2\pi \times 3.4 \, \mathrm{MHz}$ ($Q_\mathrm{c} = 2150$), respectively, the pure dephasing rate $\Gamma_\phi = 2\pi \times 2.0 \,\mathrm{MHz}$, and the Rabi frequency $\Omega_\mathrm{R}$ (bottom right). Since the Rabi frequency is proportional to the incident drive amplitude $\Omega_\mathrm{R} \propto \sqrt{P_\mathrm{in}}$, we can calibrate the attenuation $A = 90\,\mathrm{dB}$ between room temperature and our sample at the given transition frequency from the measured Rabi frequency (bottom right panel). From similar measurements at different flux points, we can infer the transfer function of our input line as a function of frequency: an additional useful feature of our double SQUID magnetometer \cite{Hoenigl2020,Lu2021}.

\begin{figure}[!t]
\begin{center}
\includegraphics[width = \columnwidth]{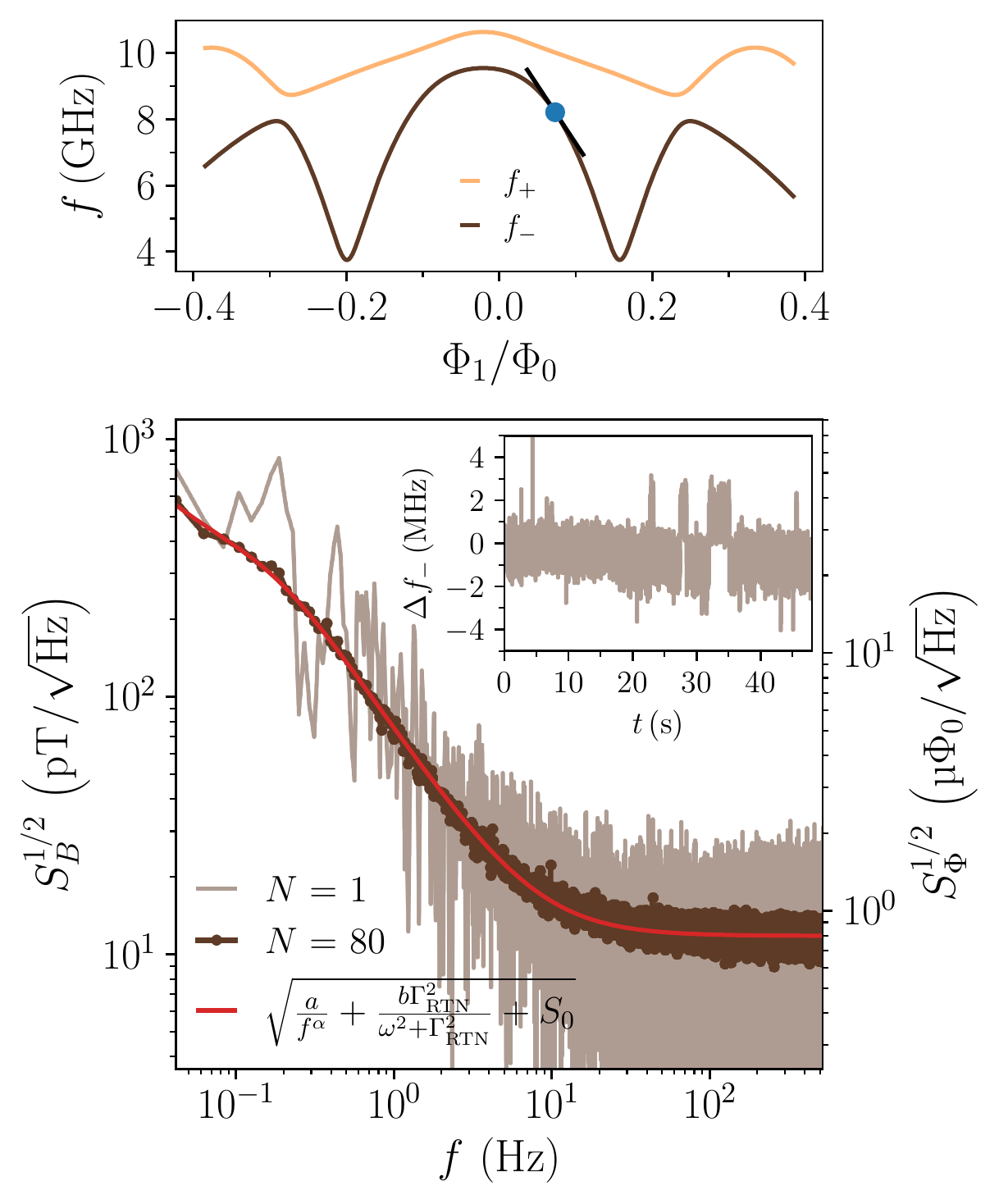}
\caption{\textbf{Noise equivalent magnetic field.} Frequency spectrum of the noise equivalent magnetic flux and magnetic field (NEF) $S^{1/2}_\Phi$ and $S^{1/2}_\mathrm{B}$, respectively, measured by monitoring time fluctuations in the transition frequency $f_-$. The data is extracted from a single time trace ($N = 1$, light brown curve), and from the average over $N = 80$ individual time traces (dark brown markers). The static magnetic flux bias $\Phi_1 / \Phi_0= 0.073$ and the drive frequency $f_\mathrm{d} = 8.8668 \, \mathrm{GHz}$ are indicated by the blue dot in the top panel, together with the linearized responsivity $\Re_\Phi$ (black solid line). At noise frequencies below $10\,\mathrm{Hz}$, the NEF spectrum shows a $1/f^\alpha$ dependence, with $\alpha = 1.42$, and the intimation of a Lorentz distribution caused by telegraphic noise with a switching rate $\Gamma_\mathrm{RTN} = 1.07\,\mathrm{Hz}$. At higher frequencies, the NEF saturates at $0.8 \, \si{\micro}\Phi_0 / \sqrt{\si{\hertz}}$ or equivalently $ 11 \, \mathrm{pT / \sqrt{Hz}}$ (white noise). The inset depicts the time trace of the resonance frequency $f_-$ corresponding to the light brown NEF, showcasing telegraphic noise.}
\label{fig4}
\end{center}
\end{figure} 

\subsection{Noise equivalent magnetic field}
\label{SEC:noise_equivalent_field}

Besides the modulation pattern caused by a static external magnetic field discussed in the previous section, an important figure of merit for every magnetic field sensor is the noise equivalent magnetic field (NEF) $S^{1/2}_\mathrm{B}$, which is a measure of the device's detection sensitivity. In general the NEF is a combination of the detectors susceptibility to magnetic fields, denoted responsivity $\Re_B$, and the possibly frequency dependent magnetic field noise amplitude $\mathcal {A}(f)$ created by fluctuations in the environment:
\begin{equation}
S^{1/2}_\mathrm{B} (f) = \frac{\mathcal {A}(f)}{\Re_B},
\end{equation}
For magnetic field sensor based on a dc SQUID, the noise equivalent magnetic field can be calculated from the noise equivalent magnetic flux using the loop area: ${S^{1/2}_\mathrm{B} = S^{1/2}_\mathrm{\Phi} A}$.

In our device, the presence of magnetic field noise causes fluctuations in the resonance frequencies over time $f_\pm \rightarrow f_\pm(t)$. Therefore, in order to measure these fluctuations, we monitor the complex reflection coefficient $S_{11}$ close to a device resonance at a constant drive frequency $f_\mathrm{d}$ and power $P_\mathrm{in}$, for a duration $T$ and a finite time resolution $\Delta t$ determined by the inverse of the intermediate frequency (IF) bandwidth. Because of the rapid phase and amplitude change with frequency around each resonance, as illustrated in Fig.\,\ref{fig3_fluo}, a change in the resonance frequency is reflected in a change of the measured reflection coefficient. Notably, in comparison to a linear detector, the power dependence of the reflection coefficient causes a non-monotonic behavior of the signal-to-noise ratio with increasing signal power. In post processing, the temporal transition frequency $f_- (t)$ is calculated for every time increment $t$ by finding the frequency detuning $\Delta / (2 \pi) = f_\mathrm{d} - f_- (t)$ between drive and transition frequency, that minimizes the geometric distance in the complex plane between the prediction according to Eq.\,\ref{EQ:reflection_qubit} and the experimental data (see App.\,\ref{ASEC:NEF}). From every time trace $f_-(t)$ recorded, first the noise equivalent magnetic flux $S^{1/2}_\mathrm{\Phi}$ is calculated as the normalized discrete Fourier transform $\mathcal{F}\{f_- (t) \}$, divided by the responsivity $\Re_\mathrm{\Phi}  = \partial f_- / \partial \Phi_1$ at the given static  ${\mathrm{flux\,\,bias}\,\,\Phi_1}$
\begin{equation}
S^{1/2}_\mathrm{\Phi} (f) =  \frac{\mathcal{F}\{f_- (t) \}}{\sqrt{\mathrm{BW}} \Re_\mathrm{\Phi} },
\end{equation} 
where the bandwidth $\mathrm{BW} = 1 / (2 T)$ is determined by the duration of the measurement.   

Figure\,\ref{fig4} depicts the noise equivalent magnetic field as a function of noise frequency $f$, measured for the lower frequency mode $f_-$. The static flux bias $\Phi_1 / \Phi_0= 0.073$ and the drive frequency $f_\mathrm{d} = 8.8668 \, \mathrm{GHz}$ are indicated in the top panel, together with the previously deduced static flux dependence of the transition frequencies $f_-$ (dark brown) and $f_+$ (light brown) already shown in Fig.\,\ref{fig3}. Due to the flux bias point, the loop area ratio can be well approximated by the larger loop area ($A = A_2$). The duration of each time trace is $T = 48\,\mathrm{s}$ measured with an IF bandwidth of $1\,\mathrm{kHz}$. The light brown curve indicates the NEF calculated from a single time trace, while the dark brown curve represents the arithmetic mean over $N = 80$ individual traces. We fit the observed frequency dependence with
\begin{equation}
S^{1/2}_\mathrm{B} (f) = \sqrt{\frac{a}{f^\alpha} + \frac{b \Gamma_\mathrm{RTN}^2}{\omega^2 + \Gamma_\mathrm{RTN}^2} + S_{0}},
\label{EQ:noise_spectrum}
\end{equation}
where the first term accounts for $1/f$-noise, while the second and third term represent random telegraphic noise (RTN) and frequency independent (white) noise, respectively. From a fit to the data using Eq.\,\ref{EQ:noise_spectrum}, indicated by the red solid line, we extract a mean switching rate $\Gamma_\mathrm{RTN} = 1.07\,\mathrm{Hz}$, potentially caused by two-level fluctuators residing inside the JJ barrier \cite{Schloer19} (see App.\,\ref{ASEC:EDoF}), and a white noise amplitude of $0.8 \, \si{\micro}\Phi_0 / \sqrt{\si{\hertz}}$ or equivalently $ 11 \, \mathrm{pT / \sqrt{Hz}}$, comparable to state-of-the-art implementations \cite{Bal2012,Jabdaraghi2017,Mart_nez_P_rez_2016,Schmelz_2017,Woelbing_2013}. The inset depicts a typical time trace, corresponding to the light brown NEF, showcasing the telegraphic noise.
\section{Conclusion}
\label{SEC:conclusion}
In summary, we have demonstrated conceptually that, by using the critical current flux modulation of two dc SQUIDs with non-integer loop area ratio, the modulation period $M$ of the combined response is enhanced substantially compared to the individual modulation periods determined by the magnetic flux quantum, which extends the unambiguously distinguishable field range significantly. Furthermore, we have demonstrated a first microwave implementation of this concept by embedding two SQUIDs into a tank circuit using conventional superconducting thin film technology. From the flux modulation of the two lowest frequency eigenmodes of our device, we find a unique field response for each applied magnetic field within $M$. The experimentally demonstrated lower bound on the modulation period is $M \geq 15\,\Phi_0$, with a theoretical maximum of $M > 625 \Phi_0$ predicted by our model, but limited to $M \approx 25 \Phi_0$ by the critical field of the tank circuit. Additionally, due to the symmetry of the modulation pattern around absolute zero-field, we are able to deduce a magnetic offset field $B_{\perp,0} = 22\,\mathrm{nT}$. The noise equivalent magnetic field measured with our device saturates at $ 11 \, \mathrm{pT / \sqrt{Hz}}$ above $100\,\mathrm{Hz}$.

By substituting the pure aluminum thin films with other superconducting materials with lager critical magnetic field, for instance granular aluminum \cite{Cohen68,Friedrich19,Winkel_2020} or Niobium compounds \cite{Samkharadze16,Shearrow18,Niepce19}, we are convinced that our concept can be extended to significantly larger magnetic fields. Furthermore, the measurement time could be improved significantly by using a circuit with smaller anharmonicity, able to sustain larger drive powers \cite{Verney19,Valenti19}. 

We are grateful to F. Valenti, J. Brehm, L. Grünhaupt and A. Stehli for fruitful discussions, and we acknowledge technical support from S. Diewald, A. Lukashenko and L. Radtke. Funding is provided by the Alexander von Humboldt foundation in the framework of a Sofja Kovalevskaja award endowed by the German Federal Ministry of Education and Research, and by the European Union's Horizon 2020 programme under Nº 899561 (AVaQus). S.G., P.W., K.B., D.R. and W.W. acknowledge support from the European Research Council advanced grant MoQuOS (N. 741276). A.V.U. acknowledges support from the Russian Science Foundation, project No. 21-72-30026. Facilities use was supported by the  KIT  Nanostructure  Service  Laboratory  (NSL). We acknowledge qKit for providing a convenient measurement software framework.

\bibliography{SQUID_magnetometer}


\appendix

\renewcommand{\appendixname}{Appendix}

\section{Device eigenmodes}
\label{ASEC:Device eigenmodes}
In order to extract the relevant device parameters from the measured transition frequencies $f_+$ and $f_-$, shown in Fig.\,\ref{fig3}, we map our device onto an effective circuit model, which we derive in the following sections. We start from a general, nonlinear model that contains the JJs as circuit elements, which we linearize to simplify the calculation. Since we are primarily interested in the modulation pattern of the eigenfrequencies with external magnetic field, a classical circuit analysis is sufficient. The qubit dynamics that we observe in our experiment (see Fig.\,\ref{fig3_fluo}), are not essential for the modulation pattern, and are discussed in a separate section.
 
\subsection{Effective circuit model}
\label{ASEC:Circuit_model}
\begin{figure*}[!t]
\begin{center}
\includegraphics[width = 2\columnwidth]{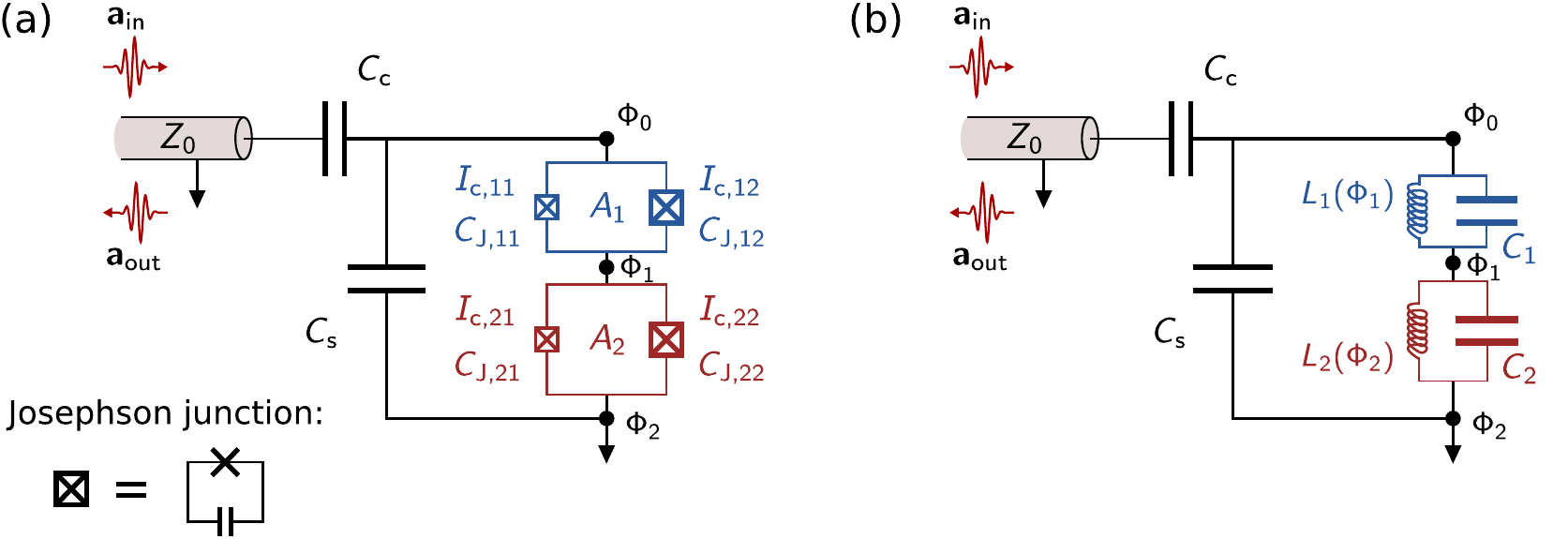}
\caption{\textbf{Circuit model.} \textbf{a)} Effective circuit model describing the magnetometer and the waveguide sample holder, both shown in Fig.\,\ref{fig2}. The magnetometer circuit consists of two dc SQUIDs, colored in blue and red, that are capacitively coupled by a capacitance $C_\mathrm{s}$. Each dc SQUID is asymmetric in terms of the critical currents of its JJs, denoted $I_{\mathrm{c},i1}$ and $I_{\mathrm{c},i2}$ with SQUID index $i$, as well as the corresponding junction capacitances, denoted $C_{\mathrm{J},i1}$ and $C_{\mathrm{J},i2}$. Notably, due to the large junction capacitances, each SQUID is a low anharmonicity transmon qubit\cite{Koch07} with flux-tunable Josephson energy $E_\mathrm{J}$. The loop areas of the SQUIDs are denoted $A_i$, with $A_1 / A_2 \approx 2.8$ in our design. The magnetometer circuit is capacitively coupled to a semi-infinite transmission line with characteristic impedance $Z_0$, representing the waveguide sample holder, via the coupling capacitance $C_\mathrm{c}$. The incident and outgoing fields are indicated by red arrows. \textbf{b)} The circuit shown in a) is linearized by considering the linear inductance of each SQUID $L_i (\Phi_i)$ only, which is a function of the flux $\Phi_i$ enclosed in the corresponding  SQUID loop, while neglecting the nonlinear contributions. The total SQUID capacitance $C_i$ is the sum of its junction capacitances, with $C_i = C_{\mathrm{J},i1} + C_{\mathrm{J},i2}$. }
\label{fig_circuit_model}
\end{center}
\end{figure*}

The nonlinear circuit model, shown in Fig.\,\ref{fig_circuit_model}a, consists of two dc SQUIDs for which we neglect the geometric loop inductance, colored in blue and red, that are shunted by a shared capacitance $C_\mathrm{s}$, representing the capacitance arising between the antenna pads (see Fig.\,\ref{fig2}b). Each dc SQUID is asymmetric in terms of the critical currents of its JJs, as well as the corresponding junction capacitances, denoted $I_\mathrm{c,ij}$ and $C_\mathrm{J,ij}$, respectively. This asymmetry is emphasized in Fig.\,\ref{fig_circuit_model}a by the different size of the circuit element symbols (box with a cross). The indices $i \in \{1,2\}$ and $j \in \{1,2\}$ indicate the SQUID and the junction number, respectively. The SQUID loop area is denoted $A_i$, and is different for both SQUIDs ($A_1 \neq A_2$, see Fig.\,\ref{fig2}c). The device is capacitively coupled to a semi-infinite transmission line with characteristic impedance $Z_0$ via the capacitance $C_\mathrm{c}$. The transmission line serves as input port for the microwave signals, and represents the copper waveguide sample holder. 

According to Eq.\,\ref{EQ:Ic_SQUID} and Eq.\,\ref{EQ:LJ_SQUID}, we model each SQUID with a magnetic flux dependent critical current and linearized kinetic inductance, respectively. For both SQUIDs we assume an individual critical current asymmetry parameter ${d_i = |I_{\mathrm{c},i1} - I_{\mathrm{c},i2}| / (I_{\mathrm{c},i1} + I_{\mathrm{c},i2})}$ and zero-field inductance. The SQUID capacitance $C_i$ is given by the sum of the junction capacitances, with ${C_i = C_{\mathrm{J},i1} + C_{\mathrm{J},i2}}$. The linearized circuit model is shown in Fig.\,\ref{fig_circuit_model}b. 

\subsection{Eigenfrequencies}
\label{ASEC:Eigen_frequencies}
In the following section, we derive analytical expressions for the eigenfrequencies of our linear circuit model (see Fig.\,\ref{fig_circuit_model}b) by finding the roots of the input impedance $Z_\mathrm{in} (\omega)$ seen by the input port. Since our model contains reactive elements only, it is these points in frequency at which the capacitive and inductive reactances cancel out, which is the general condition for resonance in an electrical circuit. The input impedance is  
\begin{equation}
Z_\mathrm{in} =  \frac{1}{j \omega C_\mathrm{c}} + \frac{1}{j \omega C_\mathrm{s} + \frac{1}{Z_\mathrm{sq}}},
\label{EQ:Zin}
\end{equation}
where $Z_\mathrm{sq}$ is the impedance of the two SQUIDs connected in series
\begin{equation}
Z_\mathrm{sq} = \frac{1}{j \omega C_1 + \frac{1}{j \omega L_1}} + \frac{1}{j \omega C_2 + \frac{1}{j \omega L_2}}.
\label{EQ:Z_SQUID}
\end{equation} 
By inserting Eq.\,\ref{EQ:Z_SQUID} into Eq.\,\ref{EQ:Zin}, and solving for the roots of the input impedance $Z_\mathrm{in}$, we find the positive eigenfrequencies $\omega_\pm$:
\begin{equation}
\omega_{\pm}^2 = \frac{\Omega_1^2 + \Omega_2^2}{2 \beta} \pm \frac{\sqrt{\Omega_1^4 + \Omega_2^4 + 2 \Omega_1^2 \Omega_2^2 (1 - 2 \beta)}}{2 \beta}.
\label{EQ:Omega_pm}
\end{equation} 
Here, the frequencies $\Omega_i$ are the bare eigenfrequencies of the SQUIDs including additional capacitive contributions from $C_\mathrm{s}$ and $C_\mathrm{c}$:
\begin{equation}
\Omega_i^2 = \frac{1}{L_i (C_i + C_\mathrm{s} + C_\mathrm{c})}.
\end{equation}
For the numerical factor $\beta$, we find
\begin{equation}
\beta = 1 - \frac{(C_\mathrm{s} + C_\mathrm{c})^2}{(C_1 + C_\mathrm{s} + C_\mathrm{c})(C_2 + C_\mathrm{s} + C_\mathrm{c})}.
\end{equation}
\subsection{Eigenmodes}
If the bare eigenfrequencies $\Omega_1$ and $\Omega_2$ are far detuned ($\Omega_1 / \Omega_2 \ll 1$ or $\Omega_1 / \Omega_2 \gg 1$), there is a small difference only between the dressed frequencies $\omega_\pm$ and $\Omega_{1,2}$, meaning the two SQUIDs oscillate almost independently. In the resonant case ($\Omega_1 = \Omega_2$), the two SQUID modes hybridize to a symmetric (+) and an antisymmetric (-) superposition of the bare modes. For the symmetric mode, the electrodes of the shunt capacitance $C_\mathrm{s}$ charge equally, while the small island between the SQUIDs charges with opposite sign. Since the modes couple to the waveguide via the dipole moment of this shunt capacitor, the symmetric mode becomes a dark mode and, thus is not visible in the resonant case. In the spectrum shown in Fig.\,\ref{fig3}, these regions are indicated by the absence of measurement data in the upper mode $f_+$. For the antisymmetric mode, the capacitor plates charge with opposite sign while the small island between the SQUIDs remains uncharged.

\section{Circuit anharmonicity}
\label{ASEC:Transmon_anharmonicity}

In our circuit implementation, the two SQUIDs are located deep in the so-called transmon regime, meaning that for both the Josephson energy $E_{\mathrm{J},i} = \Phi_0 I_{\mathrm{c},i} / (2 \pi)$ is significantly larger than the charging energy $E_{\mathrm{c},i} = e^2 / (2 C_i)$ \cite{Koch07}. In this regime, both modes are well described by a weakly non-linear Hamiltonian in the photon number basis: 
\begin{equation}
\hat{H} = \left(\sqrt{8 E_\mathrm{J} E_\mathrm{c}} - E_\mathrm{c} \right) \left(\hat{a}^\dagger \hat{a} + \frac{1}{2} \right) - \frac{E_\mathrm{c}}{2} \hat{a}^\dagger\hat{a}^\dagger\hat{a}\hat{a}, 
\end{equation}
where $\hat{a}^\dagger$ and $\hat{a}$ are the single-mode field amplitude creation and annihilation operators, respectively. The first part describes a harmonic oscillator with bare transition frequency ${\omega_0 = \frac{1}{\hbar} \sqrt{8 E_\mathrm{J} E_\mathrm{c}} - E_\mathrm{c}}$, while the second part is a non-linear term that causes the resonance frequency of the circuit to be dependent on the mean number of photons circulating in the system, usually referred to as the Kerr-term in quantum optics. The corresponding coefficient is determined by the charging energy, and determines the anharmonicity of the qubit.

In our device, the charging energies that determine the voltage fluctuations across the JJs are given by two effective capacitances $C_1^\prime$ and $C_2^\prime$, that are combinations of the SQUID capacitances $C_1$ and $C_2$, as well as the shunt capacitance $C_\mathrm{s}$ and the coupling capacitance $C_\mathrm{c}$ to the input port. From the circuit diagram shown in Fig.\,\ref{fig_circuit_model}a, we derive the capacitance matrix of our device and find the effective capacitances 
\begin{equation}
C_1^\prime = \frac{C_\star^2}{C_2 + C_\mathrm{s} + C_\mathrm{c}}
\label{EQ:C1prime}
\end{equation}
and 
\begin{equation}
C_2^\prime = \frac{C_\star^2}{C_1 + C_\mathrm{s} + C_\mathrm{c}}.
\label{EQ:C2prime}
\end{equation}
Here, $C_\star^2$ is the determinant of the capacitance matrix
\begin{equation}
C_\star^2 = C_1 C_2 + (C_1 + C_2) (C_\mathrm{s} + C_\mathrm{c}).
\end{equation} 
By inserting the circuit parameters extracted from our fit (see Tab.\,\ref{Tab:sampleoverview}) into  Eq.\,\ref{EQ:C1prime} and Eq.\,\ref{EQ:C2prime}, we find the charging energies
\begin{align*}
E_{\mathrm{c},1} / h &= 24.6\,\mathrm{MHz} \\
E_{\mathrm{c},2} / h &= 24.7\,\mathrm{MHz},
\end{align*}
which are much smaller than the Josephson energies in zero-field
\begin{align*}
E_{\mathrm{J},1} / h &= 507\,\mathrm{GHz} \\
E_{\mathrm{J},2} / h &= 504\,\mathrm{GHz}, 
\end{align*}
hence confirming our approximation. 

\section{Magnetic flux modulation}
\label{ASEC:Magnetic_flux_modulation}
Besides the zero-field circuit parameters that enter our model, which are discussed in Sec.\,\ref{ASEC:Device eigenmodes}, the frequency modulation with magnetic flux is the key feature of our device. For that reason, we relate in this section the magnetic flux enclosed in the SQUID loops to the experimental accessible quantity: the bias current $I_\mathrm{b}$ applied to the superconducting field coil.  

\subsection*{Bias current to magnetic field conversion}
\label{ASEC:bias_current_conversion}
The magnetic flux $\Phi_i$ enclosed in each SQUID depends on the loop area $A_i$ and the external magnetic field $B_\perp$ perpendicular to the SQUID plane, which we assume to be identical for both SQUIDs. This assumption is supported by the dimension of the field coil ($6\,\mathrm{cm}$ in diameter) in comparison to the SQUID area ($50 \times 50\,\si{\micro\metre^2}$). The experimental setup, including the superconducting field coil, is described in more detail in Appendix A of Ref.\,\cite{Winkel_2020}. 

Between the bias current $I_\mathrm{b}$ and the magnetic field $B_\perp$, we assume a linear relation with conversion factor $b$
\begin{equation}
B_\perp = b I_\mathrm{b}.
\end{equation}
In addition to the intentional bias field, we consider a static offset field created by the environment of the sample. We capture this offset field by an additional, static offset current $I_0$ threading the field coil, meaning that we assume an offset field identical for both SQUIDs. Hence, the magnetic flux is
\begin{equation}
\Phi_i = b \left( I_\mathrm{b} - I_0 \right) A_i.
\label{EQ:Phi_i}
\end{equation}
Similarly, we can translate the magnetic flux quantum which determines the modulation period into a modulation bias current $I_{\mathrm{p},i}$:
\begin{equation}
\Phi_{0} = b I_{\mathrm{p},i} A_i.
\label{EQ:Phi_0i}
\end{equation}
By inserting Eq.\,\ref{EQ:Phi_i} and Eq.\,\ref{EQ:Phi_0i} into Eq.\,\ref{EQ:Ic_SQUID}, we substitute the predicted magnetic flux modulation of the critical current with a bias current modulation:
\begin{equation}
\frac{\Phi_i}{\Phi_0} = \frac{I_\mathrm{b} - I_0}{I_{\mathrm{p},i}}
\end{equation}

\subsection*{SQUID loop area ratio}
The key feature of our device is the engineering of an effective modulation period by combining the modulation of two SQUIDs with different loop sizes. Since we assume a homogeneous magnetic bias field, we find a linear relation between the magnetic flux enclosed in the SQUID loops.
\begin{equation}
\Phi_2 = r \Phi_1.
\end{equation}
Here, $r = A_2 / A_1$ is the loop area ratio. Under this assumption, we find
\begin{equation}
I_{\mathrm{p},2} = I_{\mathrm{p},1} / r.
\end{equation}
Importantly, only the relative size $r$ enters our model and determines the effective modulation period. The absolute size of the loop areas $A_1$ and $A_2$ are not of importance for the modulation, but influence the device susceptibility to flux noise and magnetic field gradients.    
\subsection*{Superconducting gap suppression}
In our experiment, we observe a reduction in the frequency modulation amplitude with increasing magnetic field. Since our device is based on pure Al thin films, we address this finding to a suppression of the superconducting gap parameter $\Delta_\mathrm{Al}$ with increasing magnetic field. We derive the field dependence of the gap parameter from a two-fluid-model \cite{Tin04}
\begin{equation}
\Delta_\mathrm{Al}(B, T = 0) = \Delta_{00} \sqrt{\frac{1 - (B/B_\mathrm{c})^2}{1 + (B/B_\mathrm{c})^2}},
\end{equation} 
where the two fluids are the superconducting condensate with zero-field gap parameter $\Delta_{00} = 1.74 k_\mathrm{B} T_\mathrm{c}$, and the normal-conducting quasiparticle excitations, both described by BCS theory \cite{BCS57}. Notably, the same reduction in frequency with magnetic field can be caused by an interference effect in the large JJs. From the magnetic field range covered in our experiment, we cannot distinguish between both effects. 
Similar to the bias current modulation period $I_\mathrm{p}$ and the offset current $I_0$, we extract an effective critical bias current $I_\mathrm{b,c}$, which we can convert into a critical field with the conversion factor $b$. 
\begin{equation}
B_\mathrm{c} = b I_\mathrm{b,c}
\end{equation}

\begin{table}[t!]
\begin{center}
\caption{Summary of the fitting parameters used for the fit shown in Fig.\,\ref{fig3} (red solid line). The first column gives parameter, the second column gives the initial value fed into the fitting routine, with an explanation for all values given in the text of Sec.\,\ref{ASEC:fitting_parameters}, the third column gives the final fit value, and the last column gives a short description.} 
\begin{tabular}{||c | c  | c | c||} 
\hline
Parameter & initial value & fit value & description \\ \hline \hline
$L_1\,\mathrm{(pH)}$ & 360 & 322 & inductance SQUID 1 \\
$L_2\,\mathrm{(pH)}$ & 360 & 324 & inductance SQUID 2 \\
$C_1\,\mathrm{(fH)}$ & 700 & 722 & capacitance SQUID 1 \\
$C_2\,\mathrm{(fH)}$ & 700 & 718 & capacitance SQUID 2 \\
$\tilde{C}_\mathrm{s}\,\mathrm{(fH)}$ & 62 & 71 & effective shunt capacitance \\
$A$ & 2.8 & 2.8048 & SQUID loop area ratio \\
$d_1$ & 0.14 & 0.149 & SQUID asymmetry 1 \\
$d_2$ & 0.14 & 0.184 & SQUID asymmetry 2 \\
$I_\mathrm{p}\,\mathrm{(mA)}$ & 0.83 & 0.782 & modulation period \\
$I_\mathrm{0}\,(\si{\nano\ampere})$ & 100 & 418 & offset current \\
$I_\mathrm{b,c}\,\mathrm{(mA)}$ & 20 & 19.20 & critical bias current \\ \hline
\end{tabular}
\end{center}
\label{Tab:sampleoverview}
\end{table}

\section{Fitting parameters}
\label{ASEC:fitting_parameters}
In this section, we give an overview on all fitting parameters entering our model, and their values extracted from the fit to the experimental data shown in Fig.\,\ref{fig3} (solid lines). In addition, we compare the extracted fit values to their initial estimates, which we deduce from additional testings, for instance SEM imaging or finite-element method simulations.

The fitting parameters are the SQUID zero-field (linear) inductances $L_1$ and $L_2$, the SQUID capacitances $C_1$ and $C_2$, the effective shunt capacitance $\tilde{C}_\mathrm{s} = C_\mathrm{s} + C_\mathrm{c}$, the SQUID critical current asymmetries $d_1$ and $d_2$, the SQUID loop area ratio $r$, the bias current modulation period $I_\mathrm{p}$, the effective offset current $I_0$, and the effective critical bias current $I_\mathrm{b,c}$. All fitting parameters are summarized in Tab.\,\ref{Tab:sampleoverview} 
\subsection*{Initial estimates}
From SEM images of samples taken from the same batch (see Fig.\,\ref{fig2}c), we extract the average JJ overlap areas for the small and the large JJs, $A_\mathrm{JJ,1} = (6 \pm 0.5)\,\si{\micro\metre^2}$ and $A_\mathrm{JJ,2} = (8 \pm 0.5)\,\si{\micro\metre^2}$, respectively. Assuming a constant junction capacitance per unit area of $c_\mathrm{JJ} = 50\,\frac{\mathrm{fF}}{\si{\micro\metre^2}}$, we deduce a mean SQUID capacitance of $C_1 = C_2 = 700\,\mathrm{fF}$. The estimate for the effective shunt capacitance $\tilde{C}_\mathrm{s} = 62\,\mathrm{fF}$ is obtained using the eigenmode solver of a commercial finite-element method simulator. The SQUID inductances $L_1 = L_2 = 350\,\mathrm{pH}$ are chosen such that the bare SQUID resonance frequencies are on the order of the measured frequencies in zero-field, which are on the order of $10.0\,\mathrm{GHz}$.

From the overlap areas of the JJs, and assuming a homogeneous critical current density $j_\mathrm{c}$, we estimate the SQUID asymmetries $d_1 = d_2 = (A_{JJ,2} - A_\mathrm{JJ,1})/(A_{JJ,2} + A_\mathrm{JJ,1}) = 0.14$. The SQUID loop areas $A_1 = 50 \, \si{\micro\metre^2}$ and $A_2 = 140 \, \si{\micro\metre^2}$ are deduced from the same SEM images as the overlap areas, resulting in an estimated loop area ratio of $A = 2.8$. The bias current modulation period $I_\mathrm{p}$, the effective offset current $I_0$, and the critical bias current $I_\mathrm{b,c}$ are estimated from the measurement data.
 
\begin{table}[t!]
\begin{center}
\caption{Summary of circuit parameters deduced from the fitting parameters extracted from the fit shown in Fig.\,\ref{fig3} (solid lines).} 
\begin{tabular}{||c | c  | c||} 
\hline
Parameter & value & description \\ \hline \hline
$\omega_\mathrm{p,1}\,\mathrm{(GHz)}$ & $2 \pi \times 10.438$ & plasma frequency 1 \\
$\omega_\mathrm{p,2}\,\mathrm{(GHz)}$ & $2 \pi \times 10.435$ & plasma frequency 2 \\ 
$b (\mathrm{mT / A})$ & 52 & conversion factor $I_\mathrm{b}$ to $B_\perp$\\
$B_\mathrm{0}\,\mathrm{(nT)}$ & 22 & magnetic offset field \\
$B_\mathrm{c}\,\mathrm{mT}$ & 1.00 & critical magnetic field \\ \hline
\end{tabular}
\end{center}
\label{Tab:sampleoverview_2}
\end{table}

\begin{figure}[!t]
\begin{center}
\includegraphics[width = 1\columnwidth]{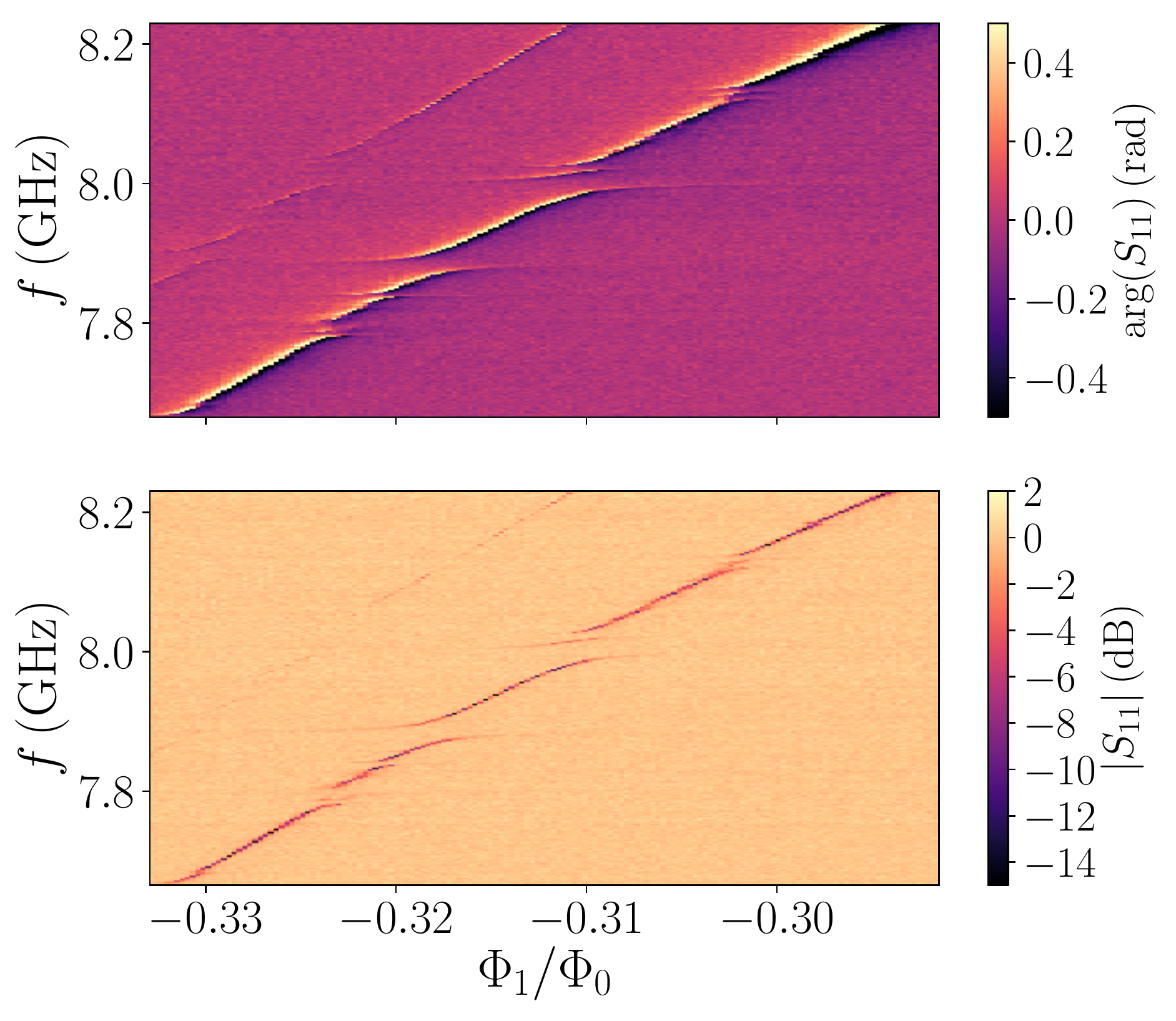}
\caption{\textbf{Environmental degrees of freedom.} Phase $\arg(S_{11})$ (top panel) in radians and amplitude $|S_{11}|$ (bottom panel) in decibel of the complex reflection coefficient, measured as a function of probe frequency $f$ and magnetic flux $\Phi_1$ inside the small SQUID loop in number of flux quanta. The sharp feature crossing the 2D plots along the diagonal is the response of the antisymmetric eigenmode of our device with eigenfrequency $f_-$. At several regions in frequency and flux, the eigenmode gets dressed by unintended environmental degrees of freedom, visible as anti-crossings. From the frequency splitting, we infer a coupling strength between these degrees of freedom and our device on the order of megahertz. The second feature in parallel to the main resonance is the response of a second device, which is visible in Fig.\,\ref{fig2}a, but not discussed in this article.}
\label{supfig_TLS}
\end{center}
\end{figure}

\subsection*{Final fitting results}

From the SQUID capacitances and linear inductances, we calculate the plasma frequencies of both SQUIDs $\omega_{\mathrm{pl},i} = (L_i C_i)^{-1/2}$:
\begin{equation}
\omega_\mathrm{pl,1} = 2 \pi \times 10.438\,\mathrm{GHz}
\end{equation}
and 
\begin{equation}
\omega_\mathrm{pl,2} = 2 \pi \times  10.435\,\mathrm{GHz}
\end{equation}
Since the JJs are fabricated in the same process, we expect similar plasma frequencies that are independent of the overlap area of the junctions, in case the critical current density is homogeneous over the wafer. 

The obtained loop area ratio $r$ and the critical current asymmetries $d_1$ and $d_2$ are in good agreement with the estimates taken from SEM images of a similar sample from the same batch.

From the bias current modulation period $I_\mathrm{p} = 0.782\,\mathrm{mA}$ in combination with the SEM estimate for the SQUID loop area $A_1 = 50\,\si{\micro\metre^2}$, we calculate the coil current to magnetic field conversion factor 
\begin{equation}
b = 52\,\frac{\mathrm{mT}}{\mathrm{A}}.
\end{equation}
Hence, we are able to convert the offset current $I_0 = 418\,\si{\nano\ampere}$ and the critical bias current $I_\mathrm{b,c} = 19.20\,\mathrm{mA}$ into a magnetic offset field $B_0 = 22\,\mathrm{nT}$ and a critical field $B_\mathrm{c,\perp} = 1.00\,\mathrm{mT}$ (out-of-plane). From the value obtained for the magnetic offset field, we conclude, that the $\si{\micro}$-metal shielding surrounding our sample, discussed in Ref.\,\cite{Gruenhaupt18} in more detail, provides a suitable measurement environment for our circuits.  The value for the critical out-of-plane field component is a factor of 10 smaller than the critical field measured for bulk aluminum \cite{Caplan65}. However, in thin film superconducting aluminum films, magnetic vortices are found to be present in much smaller fields \cite{Stan04}.

\section{Environmental degrees of freedom}
\label{ASEC:EDoF}
By sweeping the transition frequencies of our device, while measuring the reflection coefficient, we observe many frequency regions in which the device couples to environmental degrees of freedom (EDF), as shown in Fig.\,\ref{supfig_TLS} The coupling between both systems becomes visible due to a splitting of the resonance frequency into two or even three distinct transitions, so-called anti-crossings or avoided-level-crossings. The splitting indicates, that the coupling between the subsystems is transverse, and is potentially mediated via the electric field of the circuit and the dipole moment of the environmental degree of freedom \cite{Grabovskij12,Lisenfeld2019}. The exact origin of these degrees of freedom is not yet fully understand, with potential candidates ranging from adsorbents, to resist residuals from the fabrication process, to trapped quasi-particles.

In our case, we suspect that the majority of these EDF are located inside the barriers of our JJs, since they couple strongly to our circuit ($g \propto \mathrm{MHz}$), and the electric field is strongest between the junction electrodes. Additionally, the EDF do not couple to the waveguide sample holder, which can be inferred from the vanishing linewidth far away from the crossing. Notably, the second feature in parallel to and above the main resonance is the response of a second device fabricated on the same sapphire chip. While the SQUID parameters are similar, the antenna pads are much smaller for this device, resulting in a weaker coupling strength between the SQUIDs and a weaker coupling (smaller linewidth) to the waveguide sample holder. 

\begin{figure}
\begin{center}
\includegraphics[width = 1\columnwidth]{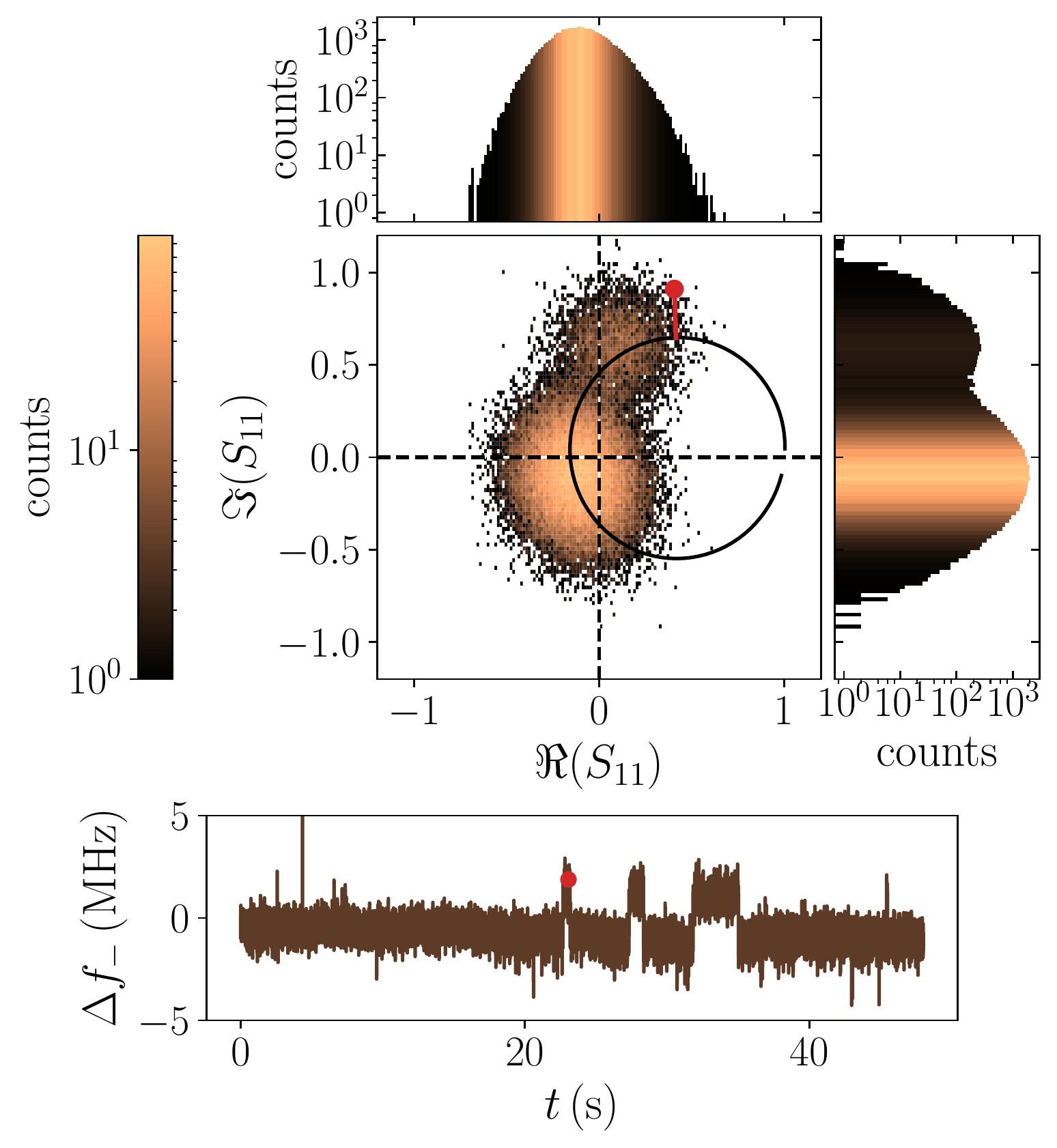}
\caption{\textbf{Reflection coefficient vs. time.} Histogram of the reflection coefficient $S_{11}$ measured at a fixed readout frequency of $f_\mathrm{d} = 8.8668\,\mathrm{GHz}$ as a function of time for a total of $T = 48\,\mathrm{s}$ every $960\,\si{\micro\second}$. The color encodes the number of counts within an area of the complex plane, as indicated by the color bar. In addition the top and right-hand panel show the projections of the whole histogram along the real and imaginary axis, respectively. The black solid line indicates the fitting result corresponding to a preceding measurement similar to the ones shown in Fig.\,\ref{fig3_fluo}, which is used to extract the change in resonance frequency $\Delta f_- = f_\mathrm{d} - f_-(t)$: As an example, the red marker highlights an arbitrary measurement outcome at time $t = t_1$, for which we find the corresponding resonance frequency at that instant of time $f_-(t_1)$ by finding the point on the solid black line the with minimal absolute distance, indicated by the red solid line. The histogram shows two distinct regions in the complex plane with enhanced probability of presence, giving rise to random telegraphic noise in the resonance frequency over time shown in the bottom panel, including the outcome for the red marker.}
\label{supfig_S11_time}
\end{center}
\end{figure}

\section{Noise equivalent field}
\label{ASEC:NEF}
As discussed in the main text, the noise equivalent field $S_B^{1/2} = S_\Phi^{1/2} A $ is derived from the noise equivalent flux $S_\Phi^{1/2}$ and the loop area $A$ of the measured SQUID. Notably, in case both modes are far detuned in frequency, meaning $f_+ \gg f_-$ or $f_- \gg f_+$, each mode can be associated to one of the SQUIDs. 
Figure\,\ref{supfig_S11_time} shows a histogram of the reflection coefficient measured as a function of time at a fixed probe frequency $f_\mathrm{d} = 8.8668\,\mathrm{GHz}$. The duration of the measurement is $T = 48\,\mathrm{s}$ with a total of 50000 points, resulting in a time resolution of $960\,\si{\micro\second}$. Prior to such a measurement, we perform a frequency resolved measurement of the reflection coefficient similar to the ones shown in Fig.\,\ref{fig3_fluo}, from which we extract the black solid line shown in Fig.\,\ref{supfig_S11_time} by fitting the data according to Eq.\,\ref{EQ:reflection_qubit}. The histogram shows to distinct areas with higher probability of presence, from which the brighter (more probable) one appears to be squeezed along the phase direction. For every reflection coefficient $S_{11}(t_1)$ measured at an instant of time $t_1$, we deduce the corresponding resonance frequency $f_-(t_1)$ of our device by finding the point on the black solid line with the closest absolute distance, as shown by the red marker and the red solid line. The two distinct states observed in the histogram give rise to random telegraphic noise in the resonance frequency versus time (bottom panel). The red marker indicates the value found for the resonance frequency for the reflection coefficient indicated in the histogram.   

From the extracted time variation of the resonance frequency, we deduced the corresponding frequency spectrum by performing a discrete Fourier transformation numerically. 
\begin{equation}
\mathcal{F}\{f_- (t) \} = \frac{1}{N_t} \sum_{n = 0}^{N_t} f_-(t_n) e^{- i 2 \pi f N_t / n }
\end{equation} 
Notably, the discrete Fourier transformation of a time trace containing $N_t$ points needs to be normalized accordingly.

\end{document}